IAC-24,E2,3-GTS.4,12,x86872

# Lunar Subterra: a Self-Integrative Unit with an Automated Drilling System

**Anthony Sfeir[a]\*, Asya Petkova[b], Sabine Chaaya[c], Karina Chichova[d], Marta Rossi[e], Anna Vock[f], Alessandro Mosut[g], Akshayanivasini Ramasamy Saravanaraj[h], Dr. Valentina Sumini[i], Dr. Tommy Nilsson[j]**

[a] *Politecnico di Milano, Italy*, anthony.sfeir@mail.polimi.it

[b] *Politecnico di Milano, Italy*, asya.petkova@mail.polimi.it

[c] *Politecnico di Milano, Italy*, sabine.chaaya@mail.polimi.it

[d] *Politecnico di Milano, Italy*, karina.chichova@mail.polimi.it

[e] *Politecnico di Torino, Italy*, marta.rossi@polito.it

[f] *Politecnico di Milano, Italy*, annalea.vock@mail.polimi.it

[g] *Politecnico di Milano, Italy*, alessandro.mosut@mail.polimi.it

[h] *Politecnico di Milano, Italy*, akshayanivasini.ramasamy@mail.polimi.it

[i] *Politecnico di Milano, Italy*, valentina.sumini@polimi.it

[j] *European Space Agency (ESA), Germany*, tommy.nilsson@esa.int

\*Corresponding Author

## Abstract

As humans venture deeper into space, the need for a lunar settlement, housing the first group of settlers, grows steadily. By means of new technologies such as in situ resource utilisation (ISRU) as well as computational design, this goal can be implemented in present years. Providing the first arrivals with an immediate underground habitat safe from radiation and other environmental constraints is of crucial importance to initialise a prolonged mission on the Moon. The project's proposal revolves around the idea of establishing a base which provides an immediately habitable space with the possibility for future expansion. Advanced construction methods and sustainable practices lay the groundwork for a permanent human presence, predominantly based on ISRU.

The narrative outlines a two-phase initiative aimed at the foundation of the Lunar Subterra, followed by an extension of the habitat above ground. The mission initiates upon the arrival of the Lander on the lunar surface, which delivers the initial circular modules designed to respect a common spaceship payload diameter. Subsequently these units are ejected onto the ground where they integrate themselves autonomously into the lunar soil using their core's integrated drilling mechanism, immediately forming a temporary safe habitat. In the following phase, the unit inflates a Kevlar membrane on its top to create an additional habitable space above the subterranean modules, which provides a habitable space on the surface for up to 6 settlers. Utilising the local regolith soil, a protective shield is 3D-printed above the inflatable membrane. This shield is individually adapted to each module, following the use of a structural analyser software, ensuring efficient material use by outlining stresses and displacement, guaranteeing an optimised printed structure. The interior inflatable membrane dimensions are obtained using an algorithm-based computational design, which optimises the membrane's size and interior habitable space area, following parameters such as tensile strength and internal pressure. In later stages of the settlement, the core showcases its multifunctionality by acting as a vertical connector (1.5m) between under- and aboveground areas. In this manner the proposed design stands out through its rapid implementation and infinite expansion possibilities through sustainable approaches. Following our collaboration with the PoliSpace Sparc Student Association group, a Virtual Reality (VR) reproduction enabled quick iterative testing of the habitable space with the use of a Meta Quest 2 headset. This not only allowed an evaluation of the environment and its impact on human residents but also eradicated the need for tangible models to conceptualise the idea, enabling rapid user-centred design and implementation in the future of space exploration.

**Acronyms/Abbreviations**

- Computer-Aided Design (CAD)
- Galactic Cosmic Rays (GCR)
- Human Landing System (HLS)
- In Situ Resource Utilisation (ISRU)
- Mining Automation Program (MAP)
- Real-Time 3D (RT3D)
- Solar Particle Events (SPE)
- Virtual Reality (VR)

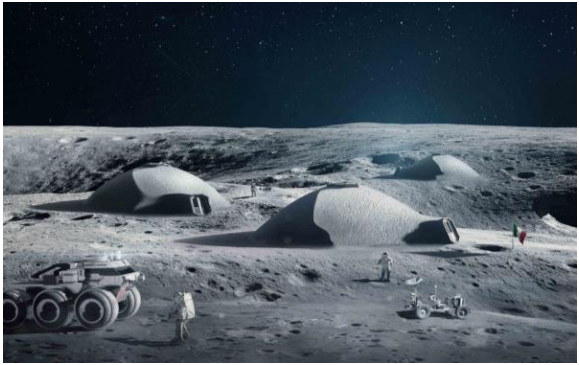

Fig. 1. Lunar Sub-Terra Overall Vision Rendering

**1. Introduction**

As humanity pushes the boundaries of space exploration, the Moon stands out as a pivotal stepping stone for future missions, including those aimed at Mars. Its proximity to Earth makes it an ideal testing ground for developing and refining technologies needed for sustainable living in the harsh conditions of space. Establishing a sustainable lunar settlement not only provides a platform for scientific research and resource extraction but also helps develop the technologies and strategies required for living in extreme, off-Earth environments.

However, establishing a lunar habitat involves overcoming significant challenges, including exposure to cosmic radiation, extreme temperature fluctuations, micrometeorite impacts, and the absence of an atmosphere. In addition, logistical issues such as transporting materials from Earth and constructing reliable and scalable habitats remain a significant obstacle.

To address these challenges, this paper introduces an innovative concept for a lunar base that can be instantly deployed and easily expanded into a larger permanent settlement. The primary base is envisioned to autonomously deploy itself on the Lunar surface, providing immediate safety and functionality for the initial group of settlers. Furthermore, the paper details the evolution of the base from an instantly deployable unit into a permanent settlement, describing how it functions and adapts over time, laying the foundation for a larger lunar colony. This approach highlights the concept's potential to play a key role as a prototype in the long-term development of lunar habitats and future space exploration.

*1.1 Lunar Frontier*

One of the primary conditions for a permanent human presence on the Moon is its proximity, predisposing rapid transportation of base facilities and astronaut teams. This logistical advantage is crucial for establishing a sustainable human presence. The relatively short travel time—approximately three days—enables quick resupply missions and emergency evacuations, making the Moon a practical option for ongoing human activity.

Despite the promising opportunities for resource utilisation on the Moon, several significant challenges must be addressed to ensure the successful establishment of a permanent settlement. One of the primary concerns is the harsh lunar environment, characterised by extreme temperature fluctuations that can range from -173 degrees Celsius during the night to 127 degrees Celsius in direct sunlight[1]. Additionally, the Moon's surface is exposed to high levels of radiation due to the lack of a protective atmosphere and magnetic field. Astronauts living on the lunar surface would face increased risks from cosmic rays and solar particle events, which implies the need for effective radiation shielding in habitat designs. Moreover, the isolation and confinement associated with living on the Moon can have psychological impacts on astronauts, making mental health support and social dynamics critical components of habitat design. The limited availability of resources and the need for self-sufficiency may also create logistical hurdles, particularly in terms of maintaining life support systems and ensuring a continuous supply of essential materials.

However, the lunar poles have been identified as suitable locations for habitation due to their unique environmental conditions. Unlike other areas on the Moon that experience extreme temperature fluctuations, the poles maintain a more stable thermal environment, with average surface temperatures around -50 degrees Celsius (-58 degrees Fahrenheit)[2]. This stability simplifies the design of habitats and reduces the energy required for thermal control, contributing to a more pleasant atmosphere for future inhabitants. Moreover, the lunar poles are not only sunlit but also in proximity to significant water ice deposits[3], particularly in permanently shadowed craters. It has been observed that these cold traps may hold enough water ice, crucial for supporting human life. Water can be used for drinking, as well as for producing oxygen and hydrogen (in theory usable for rocket fuel). The combination of constant sunlight, which simplifies energy production through solar panels, and access to water resources makes the Moon an ideal candidate for a permanent human settlement, capable of supporting both life and ongoing exploration efforts in the solar system.

*1.2 Objectives*

*1.2.1 Instant Lunar Base: the stepping stone of a permanent settlement*

Following the above mentioned prerequisites, an instant base is required upon arrival to provide the astronauts with protection from the hostile environment. This outpost should provide protection from intense cosmic radiation, solar flares, drastic temperature changes, and micrometeorites, while also meeting basic human needs for oxygen and water. Acting as a foundation of the permanent settlement, this instant base will allow the astronauts to start conducting their research from the very beginning of their mission and facilitate the further construction and expansion of the new, permanent one.

However, establishing a new habitat for humanity on the Moon involves complex tasks that necessitate specialised on-site research. Investigating the potential extraction and utilisation of in-situ materials is crucial, as it will minimise the reliance of future lunar crews on expensive material deliveries from Earth. Although the available energy sources on the Moon are known, a detailed strategy for their appropriation specific to the location is required. Additionally, both the technical and spatial qualities of the design proposal must be thoroughly examined before permanent settlement can be considered.

*1.2.2 Underground Shelter*

One of the most plausible solutions, providing natural shielding and protection from the harsh environment, is the underground shelter. Such an outpost delivers a crucial precedent for survival once the essentials for life are met[4]. In order to mitigate the intense bursts of radiation, which have an annual dosage of approximately 300 mSv on the Moon [5],it has to be executed with a sufficient depth[4]. In this way also meteorites rain and solar flare could not reach the subterranean base, which becomes a safe environment for direct exploitation. Furthermore, positioning it under the ground offers the possibility for mining the lunar regolith, enabling ISRU to facilitate future expansion of the habitat.

*1.2.3 Extension of The Habitat*

While an initial outpost provides immediate protection, it is inadequate for long-term habitation due to limited space and insufficient natural light, which can adversely affect the mental health of the inhabitants[6]. To effectively address this challenge, establishing a visual connection to the exterior environment is essential. This necessitates an extension of the outpost above ground level. A viable solution for achieving this expansion is the incorporation of an inflatable structure[7]. This structure would not only increase the available living space but also would provide functions required to support the needs of the future inhabitants. Given it will be situated above ground, additional radiation protection is necessary. The paper will explore the utilisation of lunar regolith as a natural shield against harmful radiation, leveraging its properties to enhance safety.

*1.3 Lunar regolith as a source for radiation shielding*

This section explores the potential of the lunar regolith as a primary resource for shielding permanent habitats, focusing on its role in radiation shielding and ISRU. ISRU is defined as the "extracting, transforming, storing, and utilising material resources under the specific conditions of the lunar surface environment"[5].

Lunar regolith is composed of fine dust and rocky fragments, generated through a process called gardening. The concept of "gardening" refers to the natural processes that create lunar soil over geological time. The Moon is continuously bombarded by meteoroids and micrometeoroids, which contribute to the formation of regolith. The natural properties of regolith allow it to act as an effective barrier against radiation sources, including Galactic Cosmic Rays (GCR) and Solar Particle Events (SPE) that pose risks to health. Another benefit of the material is its structural integrity: regolith can be compacted and used as a load-bearing material, providing structural support for habitats while also serving as a shield against micrometeorite impacts and extreme temperature fluctuations[8].

While regolith is abundant, its effective utilisation requires advanced technologies for extraction and processing which includes the need for automated systems to manage the construction process in the inhospitable environment. The construction methods will be highly automated due to the limited availability of human labour in the early stages of establishing the lunar settlement. This process will require reliable robotic systems for regolith processing and habitat assembly. The advancement of robotics over the past five decades has significantly transformed terrestrial mining and construction practices, establishing a foundation for similar applications on the Moon[9]. The methodologies developed for Earth-based mining were made through programs like the Mining Automation Program (MAP)[4]. Through the MAP, various tasks including exploration drilling, tunnel drilling, explosives loading and ignition, material removal and transport, rock stabilisation, and longhole drilling for ore transportation are now applicable for lunar habitat construction and can also be used for mining water and other essential resources on the Moon[4]. The adaptation of these techniques for extraterrestrial environments allows for remote operation of machinery increasing safety and speed, which is crucial given the risks associated with human presence on the lunar surface. Another vital utilisation of robotics can be water mining where the creation of a collaborative underground lunar outpost and mining facility focused on extracting water could be crucial for sustaining long-term human presence on the Moon. Through this method we can facilitate the establishment of a permanent lunar base.

*1.4 Immersive Interiors: Virtual Reality for human-centred Design Solutions*

Space habitation solutions have historically been designed with a primary focus on ensuring basic life support, prioritising the technical necessities required to sustain human life in extreme environments. Addressing immediate physiological needs of the crews - such as oxygen supply, radiation shielding, and temperature regulation - thus consistently dominated the design agenda. Comparatively less consideration was given to psychological factors related to interior design, such as comfort, aesthetics, or long-term livability[10].

With renewed interest in lunar exploration, future missions are poised to grow more ambitious in both scope and duration. Simultaneously, ongoing democratisation of access to space, fueled by the rise of the private space industry, is expanding and diversifying the user base. Against this backdrop, there is a growing recognition that future design efforts must go beyond traditional utilitarian requirements[11]. Once considered an afterthought, factors such as psychological well-being of the increasingly diverse space travellers will inevitably play a key role in the success of future long-term expeditions. This shift amplifies the need to integrate inclusive user-centred design principles into future habitat development projects, ensuring that both physiological and emotional requirements are addressed.

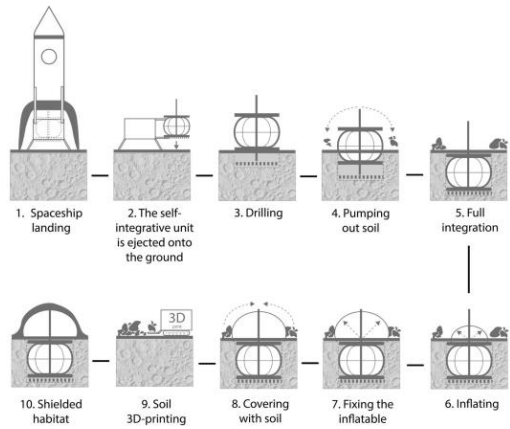

Design efforts in this area have traditionally relied on analogue field studies conducted in remote locations, such as Antarctica or the Utah desert, to approximate some of the challenges associated with off-Earth habitation[12]. While these studies are generally useful, they are often costly and only involve a small number of participants, leading to relatively slow and less inclusive design processes.

An emerging alternative in the space industry involves simulating field studies using immersive Virtual Reality (VR) technology. This method offers a cost-effective and safe way for users to experience hypothetical design solutions within representative extraterrestrial environments, thereby enabling evaluation and rapid iterative development of novel concepts in a controlled setting. Moreover, the comparatively accessible nature of VR allows for the involvement of a wide range of prospective users from the earliest stages of a design process, helping to bridge the gap between traditional space systems engineering and user-centred design approaches[11].

Drawing on these advantages, we created a digital replica of the Lunar Subterra habitat concept and utilised VR for its initial design assessment. Our goal was not only to validate interior architectural elements crucial for the long-term well-being of future occupants, but also to demonstrate VR's unique capacity to support inclusive design processes. In the remainder of this section, we offer a detailed reflection on this approach and its outcomes.

**2. Methodology**

1. Researching previous documentation on lunar regolith, in order to evaluate the integrity of the used techniques
2. Establishing a clear phasing strategy to define the parameters for the execution of each of the modules
3. Designing the self-integrative unit, outlining the adaptation to the size of the spaceship, the multifunctionality of the core and the use of materials
4. Using computational design programs to simulate the inflatable membrane via Kangaroo, Octopus and Karamba
5. Developing a shielding strategy through comparative analysis of materials for the shell
6. Demonstrating how virtual reality benefits the design through verification of the habitat's spatial performance

**3. Lunar Subterra: The habitat, theory and implementation**

*3.1 Concept Timeline*

Fig. 2. Lunar Sub-Terra Evolution

The project employs a phased construction approach. Initially, a pod designed for integration with most Lunar Landers is ejected onto the Moon's surface. It then submerges itself into the ground using an Integrated drilling mechanism in its design, creating an immediate underground base that provides initial shelter for the first group of settlers. Once they are settled, an inflatable membrane deploys on its top. Although connected, the primary module can operate independently from the inflatable extension, which is subsequently covered with excavated lunar regolith to create a more permanent habitat.

The primary pod is a prefabricated and autonomous living module. In this regard, since all essential functions required to sustain the settlement will be located within this module, it should be treated and designed as the safest and most secure space available. On the other hand, the added Inflatable should extend the functionality layer beyond the basic survival needs, it will provide crucial functions for the settlers' research and well-being—both physical and psychological—by offering additional space for social areas, a gym, etc. This extension serves as a public and interactive area necessary, though not vital in emergencies. This synergy between primary and secondary infrastructure influences significantly all further design decisions, such as those concerned with the interior layout distribution, the choice of materials, or the arrangement of needed equipment.

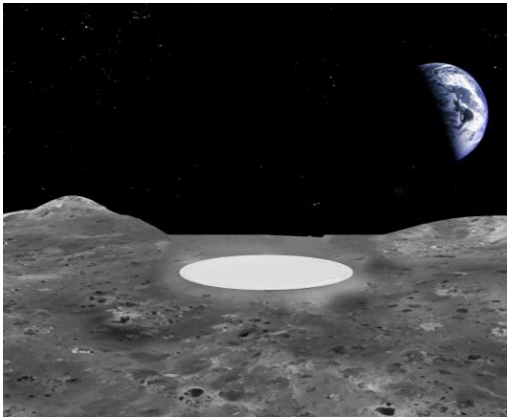

Although further expansion is not the primary objective of this paper, it is briefly discussed here. It involves linking multiple similar base modules, each adhering to the underground and inflatable concept. These modules would be interconnected via underground tunnels, forming a colony where multiple astronaut teams can operate independently until the connections are fully established. This growth strategy resembles the branched design proposed by D. Nagy, A. Bond, and O. Bannova, which features one vertical module for life support and two horizontal modules interconnected through nodes for shared utilities and science[13]. While the construction method is uniform, functional differences persist among the modules.

*3.2. The initial self-integrative unit*

Fig. 3. Early Conceptual render of the first unit and its activated drill

The initial unit is an automated module characterised by its "site independence," meaning it can be easily transported and deployed across various locations without being tied to a specific site. This unit is cylindrical with rounded edges, a structural choice optimised for handling internal pressure. It has a diameter of 6 metres and a height of 4 metres, with dimensions tailored to fit within the payload capacity of modern lunar landers, such as the Starship HLS [14]. This design enables smooth and rapid deployment of the pod from the spacecraft to the lunar surface. The bottom is equipped with an integrated circular drilling mechanism that matches the pod's diameter, allowing for autonomous subterranean deployment. At the pod's core a 1.5-metre wide rotational shaft is positioned, it is responsible for the functioning of the drill. It significantly influences the interior layout with its central location, dividing the internal space. This rotational core plays a critical role in the later stages of the habitat's expansion, showcasing its multifunctionality through adopting several different uses.

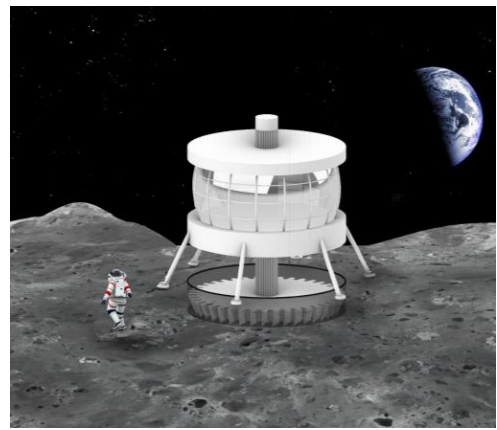

*3.3 The multifunctionality of the drilling system*

Fig. 4. Conceptual render of the first unit submerged in the lunar soil

Upon landing on the Moon, the lander ejects the cylindrical pod onto the lunar surface. Equipped with support legs, the pod elevates itself slightly to allow the integrated drill to initiate soil excavation without causing damage to the unit. The rotational core powers the drill to dig into the lunar soil, deep enough to accommodate the full height of the module. Once the excavation is complete, the rotational core facilitates the pod's descent into the ground, using a rotational motion to position it securely. This process typically takes about 20 hours[15], resulting in a habitat that is insulated from the Moon's extreme surface conditions. The materials used for the execution of the pod are resistant to the temperature fluctuations and radiation on the Moon, while its integration into the lunar soil adds an additional layer of safety. The pod's floor and ceiling are designed to accommodate all necessary mechanical, technical, and electrical systems required for the drill's operation and the overall life support of the future base.

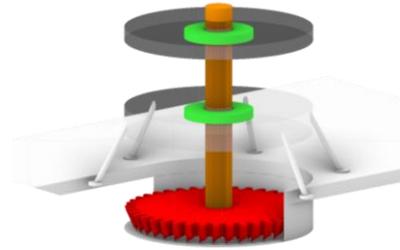

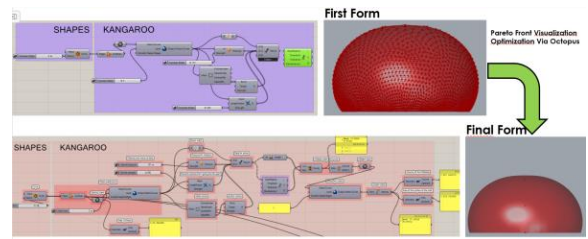

Fig. 7. Kangaroo form finding optimisation

Fig. 5. Conceptual Schematic of the Integrated Drill Mechanism

*3.4 The inflatable membrane*

Following this initial phase, the expansion of the base begins. An inflatable Kevlar membrane is automatically deployed from the top centre of the pod. This membrane, inflated by an onboard oxygenator located in the pod's ceiling, takes the shape of a dome.

A computational design approach for both the inflatable membrane and the radiation shield was used. The plug-in *Kangaroo* for the *Rhinoceros 3D* software is a Live Physics engine for interactive simulation, optimisation and form-finding directly within *Grasshopper*. It facilitated the form-finding process and defined the dimensions of the inflatable membrane, shaping the habitable upper volume while accounting for various parameters. The chosen membrane material is Kevlar[16] since it has suitable properties for inflatable membranes, such as high tensile strength, impact

Once the inflatable is established, the rotational core showcases its multi-functionality by acting as a vertical connector, linking the new upper level with the subterranean pod to form a larger settlement. The inner part of the rotational core acts as a shaft, inside of which a ladder is installed to facilitate movement from the bottom to the mezzanine level of the inflatable structure. Although the rotational core only extends to the top of the pod, the partition layout within the inflatable follows a circular pattern, as if the core continues to the top of the dome. This design ensures a seamless and efficient vertical connection throughout the entire structure.

Fig. 6. Section Drawing of the overall Lunar Sub-Terra

Despite the Kevlar membrane's protection, additional shielding is necessary to protect against external harsh conditions. A regolith shield is 3D-printed over the inflatable using the lunar soil extracted through the excavations by the integrated drill. This shield covers the entire inflatable membrane, providing protection against impacts and direct exposure to the lunar environment.

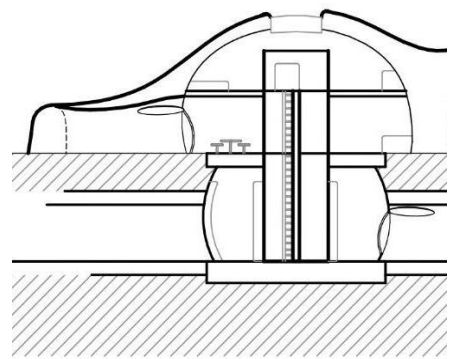

resistance, and lightweight nature. Its flexibility allows it to form tough, puncture-resistant outer layers. Kevlar also has excellent thermal stability, low thermal conductivity, and abrasion resistance, offering protection against extreme lunar temperatures and surface wear[17]. These material properties were reported into *Kangaroo* (especially tensile strength and internal pressure) in order to accurately generate the form of the mesh (membrane). *Kangaroo* generated a sphere-shaped mesh under several parameters and conditions, including environmental constraints (SPEs and GCR radiations, extreme temperature variations, microgravity and absence of atmospheric pressure) and needs (floor areas adapted to 6 crew members). The criteria used for evaluation were the pressurised volume, the kevlar shell surface and the habitable area.

The main design variable was the interior radius of the semi-sphere mesh. The first form generated by *Kangaroo* had a radius of 6 m corresponding to a total floor area of 247 sqm with 2 floor levels. Then a multi-objective optimisation was conducted via the *Octopus* plug-in in order to further improve the initial form. A Pareto Filter/ Pareto Front Visualisation was integrated to navigate the process according to 3 objectives, each represented by one axis in the diagram:

1. Maximise volume
2. Minimise shell surface (in order to reduce material and resulting costs)
3. Maximise total floor areas

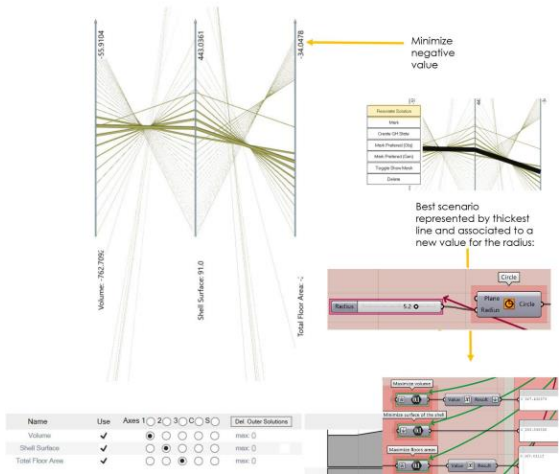

Fig. 8. Octopus multi-objective optimisation determining r=5.2m

After running *Octopus'* optimisation, a set of different solutions was obtained, outlining the best scenario for the Kevlar inflatable membrane to be a semi-sphere of a 5.2 m radius with a total floor area of 167 sqm.

Finally, the last step of the computational design flow is the structural analysis of the inflatable membrane using the *Karamba 3D* which is an interactive, parametric engineering tool that allows you to perform quick and accurate Finite Element Analysis in *Grasshopper*. Displacement, stress, horizontal and vertical isocurves were analysed to verify the structural rigidity and robustness of the obtained volume.

*3.5 The lunar regolith shield*

Following a similar process to the one described in 3.4, a radiation shield protecting the membrane is conceived using *Karamba3D*. The data from the optimised shape of the inflatable is then used to create an offset, which serves as the foundation for the regolith shell, ensuring that the membrane's surface and the outer shell do not overlap.

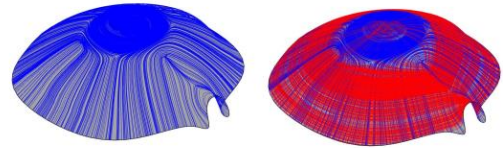

Fig. 9. Shield Horizontal and vertical Iso curves

Then via *Karamba 3D* horizontal and vertical Iso curves are determined, based on which robots will 3D print the regolith into a protective shell over the inflatable semi-sphere membrane. This process involves depositing layers of regolith in a precise, dome-like structure that conforms to the habitat's shape. The regolith is sintered, using heat or binding agents, to create a solid, durable shell [18]. The final shape is determined by stress and displacement diagrams obtained using *Karamba3D*, ensuring structural integrity based on the weight of the lunar regolith and the shape of the shield.

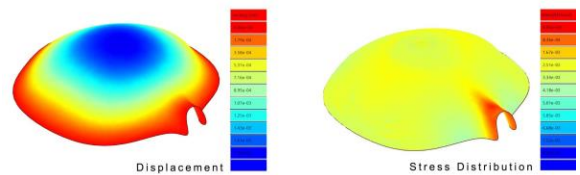

Fig. 13. Shield Displacement and Stress distribution analysis with *Karamba 3D*

*3.6 The interior distribution and design*

The obtained inflatable volume adds up to the initial underground base. The vertical connection between both is located in their coinciding centre and exploits the space formerly used by the drilling system. Each level hosts different functions, assigned according to their requirements in terms of space, sound and visual comfort.

The interior design was structured around a conceptual daily schedule that vaguely leans on the routine of an astronaut on the International Space Station: eight hours allocated for laboratory work, eight hours for sleep, two hours for physical exercise, two to three hours for meals and social interactions, and one to two hours of personal downtime. This schedule guided the spatial allocation of key functional areas, including the kitchen/social space, sleeping quarters, working facilities, and laboratory environments.

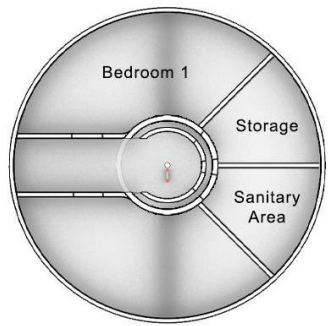

Sub Terra Floor Plan

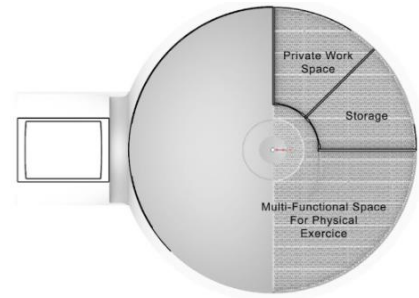

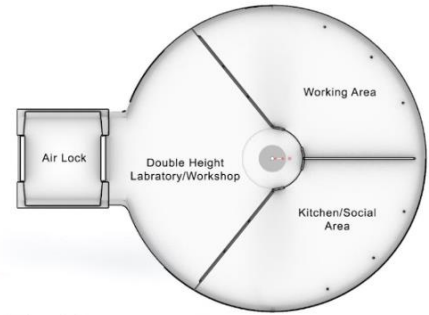

Fig. 10. Lunar Sub-Terra Ground and Mezzanine Floor Plans.

Given the design constraints imposed by the rocket payload, the underground pod was optimised for sleeping and working areas to maximise protection against radiation and micrometeorite impacts. This configuration also aimed to mitigate feelings of spatial confinement. The laboratory, positioned on the ground level, facilitates direct access to post-extravehicular activities while minimising contamination risks. The kitchen, strategically located behind the lab, serves as both a dining and social hub, recognising the fundamental role of communal eating in astronaut well-being. Adjacent to the kitchen, a working area equipped with desks and computers supports extended work periods and may include a green area for added comfort.

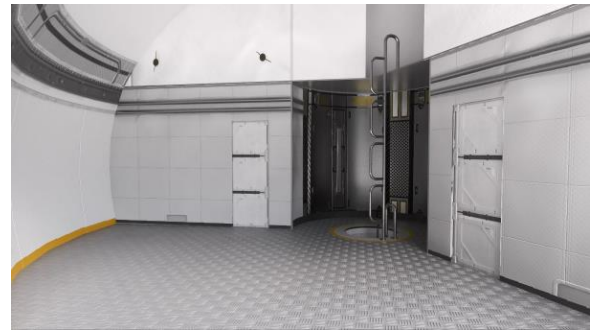

Fig. 11. Interior Render

On the upper level, supported by a metal mezzanine independent of the inflatable structure, the design incorporates a multi-purpose workstation, logistics stowage, and a multifunctional space for physical exercise. This layout balances functional efficiency with comfort, ensuring adequate facilities for both work and leisure. Additionally, a central storage area between floors houses critical life support systems essential for habitant survival.

*3.7 VR Implementation and Initial Evaluations*

Following the initial architectural design phase utilising *Rhinoceros* and *Grasshopper* software, the project transitioned into the VR Unity environment, as justified in the preceding section. For the VR implementation, *Unity* was utilised as the development RT3D platform, while the *Meta Quest 2* headset, a high-performance hardware solution, was deployed for the final execution of the virtual experience. The software suite also included *Cinema 4D* and *Substance 3D Painter*. These tools were used to optimise the 3D models for VR and to enhance texturing, to achieve high-quality visualisation. This approach facilitated an iterative exploration of interior layouts, allowing us to refine our designs based on spatial considerations and user experience insights.

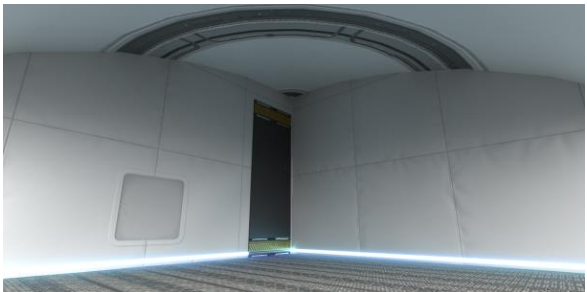

Fig. 12. Interior Render

Conceptual sketches and 3D renderings provided the team with the foundation for internal discussions and brainstorming sessions on the fundamental design and layout of the habitat. Subsequent instantiation of these concepts in an immersive VR environment allowed for a more detailed and realistic visualisation, enabling the team to assess spatial considerations and other relevant dynamics. In the formative stages of the project, before finalising the spatial design and while considering various design alternatives, we utilised virtual reality to visualise these proposals. Three students were engaged to explore the VR environments. The feedback collected from this exercise was crucial in refining and finalising the interior layout. In particular, this test revealed that ceiling height variations, compounded by the effects of microgravity, significantly influenced their perception of space, affecting overall comfort and usability assessments. For instance, environments that seemed adequately spacious on renderings and sketches were perceived as claustrophobic in VR, highlighting the need to account for both spatial and gravitational factors in the design process. Furthermore this illustrates that this technology, by providing a more human and lifelike perspective, significantly enhances the intuitive understanding and projection of spaces.

Lighting emerged as a crucial element impacting the habitability of the environment, with both general and task lighting available throughout the habitat to enhance orientation and functionality [19]. The contrast between CAD renderings and VR experiences underscored the need for precise spatial and lighting design to ensure an optimal living experience. Further detailed feedback on the final design will be gathered with more tests in the future, enabling a more comprehensive assessment of the habitat's interior.

*3.8 Future Evaluations*

The initial interior design of the habitat was concentrated on a single unit, intended to function independently with all necessary components integrated within this solitary unit. However, for future considerations, we are exploring the potential benefits of modularity within the habitat. This involves investigating how multiple units, when combined, could operate as a cohesive lunar colony. Consequently, the interior design will be adapted to meet this new challenge. Engaging with experts, including analog astronauts, professional astronauts, and VR specialists, will facilitate the process providing critical feedback

In the future, we plan to allow more individuals to interact with this environment virtually to gather additional data. This hands-on experience will be complemented by qualitative feedback gathered through post-hoc semi-structured interviews with the study participants, which will be instrumental in refining the habitat design and addressing any identified issues.

**5. Conclusion**

This paper comprehensively explores the design and implementation of the Lunar Subterra habitat, emphasising the need for innovative solutions to support long-term human settlement on the Moon. By integrating advanced methodologies such as computational design, modular architecture, and virtual reality, the project addresses both the physiological and psychological needs of future lunar inhabitants.

The phased construction approach outlined ensures a fast and secure deployment, utilising prefabricated autonomous modules and inflatable extensions to create a multifunctional living space. The emphasis on a self-integrative unit demonstrates adaptability and resilience against lunar conditions, while the use of regolith for shielding not only maximises resource efficiency but also reinforces the habitat's safety.

Incorporating virtual reality as a tool for iterative design enhances understanding of spatial dynamics and user experience, fostering a more inclusive design process. This paper emphasises the need to address psychological well-being alongside technical necessities for a sustainable space habitat development. Future considerations include examining microgravity's effects on human physiology, developing efficient ventilation systems, fire suppression systems and emergency evacuation protocols, and planning for effective food storage and waste management.

Ultimately, the Lunar Subterra project contributes valuable insights into future lunar missions, positioning itself as a potential Prototype for the next era of space Habitats.

**Acknowledgments**

We would like to thank Polispace, the student space association of Politecnico di Milano, for its support in this project.